\documentclass[a4paper]{spie}  

\newcommand{\dd}{\mathop{}\!\mathrm{d}}
 
\usepackage{amsmath,amsfonts,amssymb}
\usepackage{graphicx}
\usepackage{siunitx}
\usepackage[colorlinks=true, allcolors=blue]{hyperref}
\usepackage{comment}
\usepackage{float}
\usepackage{mathtools}

\title{On the Energy Dissipation in the Landau-Lifshitz-Gilbert Equation}

\author{Kutay Kulbak}
\author{Mohamed Iyad Boualem}
\author{Charlie Massé}
\author{Mariana Delalibera de Toledo}
\author{Vasily V. Temnov}
\affil{LSI, Ecole Polytechnique, CEA/DRF/IRAMIS, CNRS, Institut Polytechnique de Paris, F-91128, Palaiseau, France}

\authorinfo{Further author information: (Send correspondence to K.K. or V.V.T.)\\K.K. E-mail: kutay.kulbak@polytechnique.edu\\  V.V.T. E-mail: vasily.temnov@cnrs.fr}

\begin{document} 
\maketitle

\begin{abstract}
    The dynamics of magnetization near a stable equilibrium in ferromagnetic nanomagnets are examined within the Landau--Lifshitz--Gilbert (LLG) framework. For a small angle precession, the dependence of ferromagnetic resonance (FMR) frequency, the damping constant and the resulting quality factor $Q$ of the resonance on the local curvature around the free-energy minimum is systematically analyzed. Special attention is devoted to the behavior of the FMR decay time in the vicinity of bifurcation points, where the number of metastable energy minima changes and the commonly used approximation for the quality factor $Q\simeq 1/2\alpha$  (where $\alpha$ denotes the Gilbert damping) fails.
\end{abstract}


\section{Introduction}
\label{sec:intro}
Magnetization dynamics in ferromagnetic systems play a central role in modern spintronics, magnonic, and microwave technologies.  The behavior of small-angle precession around a stable equilibrium is commonly characterized by the ferromagnetic resonance (FMR) frequency, linewidth, and quality factor, quantities that determine the coherence and dissipation properties of magnetization oscillations \cite{farle1998ferromagnetic}. Within the Landau–Lifshitz–Gilbert (LLG) framework \cite{Landau1935,lakshmanan2011fascinating}, these dynamical parameters are typically expressed using simple approximate formulas that rely on the assumption of nearly circular precession and isotropic restoring forces. In particular, the widely used relation for the linewidth-to-frequency ratio (the $Q$-factor) $Q=1/2\alpha$ is routinely applied across a broad range of materials and geometries \cite{farle1998ferromagnetic, salikhov2019gilbert, BesseJMMM, ghita2023anatomy}, even though it implicitly assumes that the local curvature of the magnetic free-energy landscape is direction-independent.

However, realistic ferromagnetic structures rarely satisfy this condition. Shape, magneto-crystalline or magneto-elastic anisotropies result in elliptically polarized precession trajectories near equilibrium, governed by an energy minimum whose curvature varies along different directions on the unit sphere. Deviations from isotropy become especially pronounced near parameter regimes where the free-energy landscape undergoes qualitative changes, such as transitions from double-well to single-well configurations. In these regions, the standard expression for the damping rate loses its accuracy, and the commonly assumed value $Q=1/(2\alpha)$  can differ significantly from the textbook result.

Several exemplary attempts to treat the lifetime of magnetization precession in the presence of magnetic anisotropy have been undertaken. Analytical expressions for FMR life time in collinear (magnetization vector parallel to the external magnetic field) for a magnetic ellipsoid \cite{gurevich2020magnetization} and non-collinear configurations (magnetic field tilted with respect to the symmetry axis) \cite{alekhin2023quantitative} represent model-specific examples and do not establish a direct relationship to the ellipticity of the precession trajectory.    

In this work, the magnetization dynamics around a stable equilibrium are systematically analyzed through a unified approach based on the Hessian of the magnetic free energy density. By expressing the linearized LLG equations in a local orthonormal basis, the FMR frequency and decay rate are shown to depend exclusively on the eigenvalues of the Hessian matrix. This formulation reveals that the damping constant and quality factor depend sensitively on the curvature anisotropy. The explicit expression $$Q=\frac{1}{\alpha} \frac{\sqrt{\lambda_1\lambda_2}}{\lambda_1+\lambda_2}$$ demonstrates that the conventional approximation $Q=1/2\alpha$ is recovered only for circular energy minima \[
\lambda_1 = \lambda_2
\] Significant deviations arise whenever the energy minimum is elliptical. 

The analytical predictions are tested across several representative geometries of the magnetic shape anisotropy for a single-domain elliptical nanomagnet \cite{Osborn1945}, including a symmetric cylinder and an arbitrary ellipsoid. Numerical simulations show that the strongest departures from the usual approximation occur near bifurcation boundaries, where one of the curvatures of the free-energy landscape vanishes, and the system transitions between one and two equilibrium minima. In these regimes, the effective quality factor collapses, and the magnetization motion becomes overdamped.

Together, these results establish a general framework for predicting dissipation in ferromagnetic systems directly from the local energy landscape and demonstrate that the widely used approximation $Q=1/2\alpha$ is reliable only under restrictive geometric conditions. The Hessian-based approach developed here provides a unified description of resonance and damping behavior and clarifies the origin of deviations in anisotropic and near-bifurcation regimes. 

\section{Physical Model}

We begin by specifying the magnetic system and the assumptions underlying its description. The ferromagnet is assumed to remain uniformly magnetized such that the magnetization vector $\mathbf{M}$ has a fixed magnitude $|\mathbf{M}| = M_s$, allowing the dynamics to be parametrized solely by the angular coordinates $(\theta,\phi)$ on the unit sphere. The system is subjected to an external magnetic field $\mathbf{H}$ as well as internal effective fields originating from the crystalline anisotropies and demagnetizing effects. These contributions are incorporated through a magnetic free-energy density $F(\theta,\phi)$, from which the effective field follows the relation 
\begin{equation}
    \mathbf{H}_{\mathrm{eff}} = -\frac{1}{\mu_0M_s}\nabla_{\mathbf{m}} F, \label{eq:H_eff}
\end{equation}
where $M_s$ is the saturation magnetization. Using this expression, the dynamics of the system is governed by the Landau--Lifshitz-Gilbert (LLG) equation:
\begin{equation}
\frac{\dd\mathbf{m}}{\dd t}
= -\gamma\mu_0\, \mathbf{m}\times\mathbf{H}_{\mathrm{eff}}
+ \alpha\,\mathbf{m}\times\frac{\dd \mathbf{m}}{\dd t}, \label{eq:LLG}
\end{equation}
where $\mathbf{m}$ is the unit magnetization vector, $\gamma$ is the gyromagnetic ratio, and $\alpha$ is the Gilbert damping parameter. Within LLG framework, there exist \textit{energy extrema} positions $(\theta_0, \phi_0)$ which satisfy $\dot{\mathbf{m}}(\theta_0, \phi_0)=0$, or equivalently,
\begin{equation}
    \nabla_{\mathbf{m}} F(\theta_0,\phi_0) = 0. \label{eq:stable_eq}
\end{equation}
Among these extrema, there are \textit{energy minima} positions the magnetization gets damped into.  In the vicinity of such a minimum, the energy landscape can be approximated by its expansion, and the magnetization dynamics can be modeled as a simple damped harmonics oscillator. In order to express this small oscillation behavior, we introduce the local Cartesian system defined on the tangent space of the magnetization sphere, centered at the minimum location. We orient the Cartesian coordinate system such that its handedness aligns with the ambient spherical coordinate system, and define the axes as $(u,v)=(\delta\theta, \sin \theta_0 \delta\phi)$ .

\begin{figure}[H]
    \centering
    \includegraphics[width=0.8\linewidth]{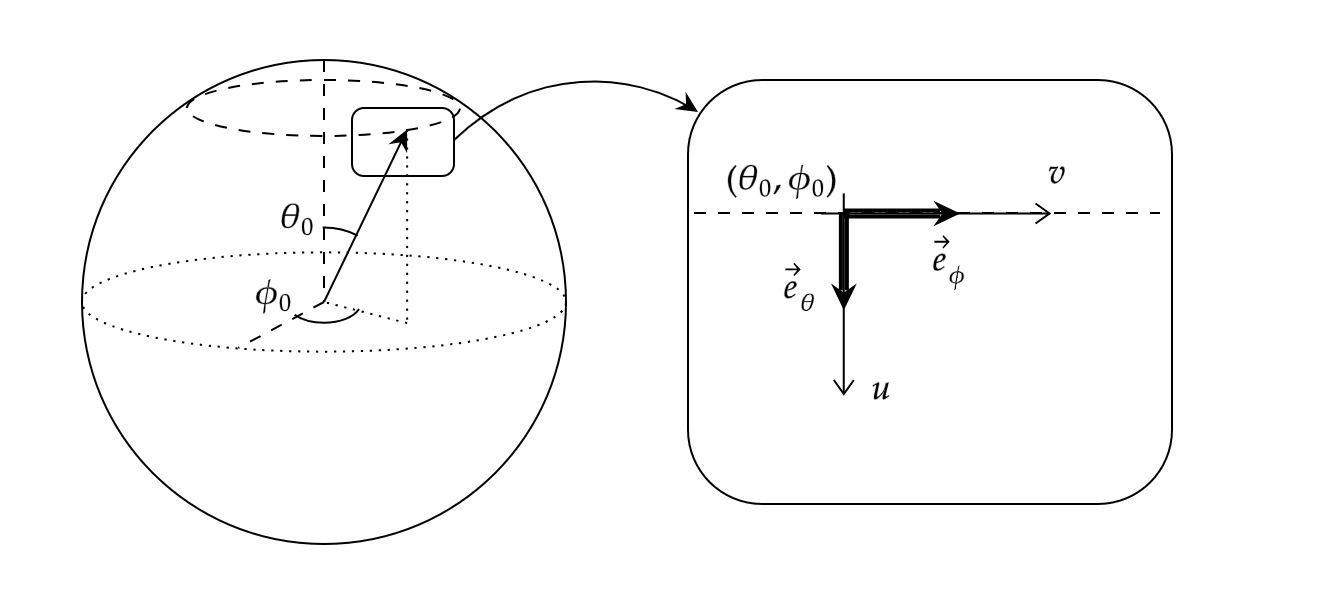}
    \caption{Visualization of the Local Cartesian Coordinate System}
    \label{fig:enter-label}
\end{figure}

\section{Theoretical Framework}

The right-hand-side of \eqref{eq:LLG} can be cast into the form
\begin{equation}
\frac{\dd\mathbf{m}}{\dd t}
= -\frac{\gamma\mu_0}{1+\alpha^2}
\big(
\mathbf{m}\times\mathbf{H}_{\mathrm{eff}}
+ \alpha\,\mathbf{m}\times
(\mathbf{m}\times\mathbf{H}_{\mathrm{eff}})
\big).
\end{equation}
The effective field described in \eqref{eq:H_eff} can be written in the spherical coordinate system as
\begin{equation}
H_{\theta} = -\frac{\partial_{\theta}F}{\mu_0M_s}, \qquad 
H_{\phi} = -\frac{\partial_{\phi}F}{\mu_0M_s\sin\theta}.
\end{equation}
Expressing $\mathbf{m}$ in spherical coordinates leads to
\begin{align*}
\dot{\theta} &= 
\frac{\gamma\mu_0}{1+\alpha^2}\,(H_\phi + \alpha H_\theta), \\
\dot{\phi}\sin\theta &= 
\frac{\gamma\mu_0}{1+\alpha^2}\,(-H_\theta + \alpha H_\phi).
\end{align*}

\subsection{Linearization around a stable equilibrium}
Let $(\theta_0,\phi_0)$ be a stable minimum of the free energy given by Eq. \eqref{eq:stable_eq}. Small deviations from equilibrium are written as
\[
u = \delta\theta, \qquad v = \sin\theta_0\,\delta\phi.
\]
The first-order expansion of the free-energy derivatives is
\[
\partial_{\theta}F = \partial_{\theta\theta}F\,\delta\theta + \partial_{\theta\phi}F\,\delta\phi,
\qquad
\partial_{\phi}F = \partial_{\theta\phi}F\,\delta\theta + \partial_{\phi\phi}F\,\delta\phi.
\]
Rearranging the LLG equation in terms of the local Cartesian coordinates gives
\begin{align*}
    \dot{u} &= \beta\left[ \left( -\alpha \partial_{\theta\theta}F - \frac{\partial_{\theta\phi}F}{\sin\theta_0} \right) u + \left( -\frac{\alpha \partial_{\theta\phi}F}{\sin\theta_0} - \frac{\partial_{\phi\phi}F}{\sin^2\theta_0} \right) v \right],\\
    \dot{v} &= \beta\left[ \left( \frac{\alpha \partial_{\theta\phi}F}{\sin\theta_0} + \partial_{\theta\theta}F \right) u + \left( -\frac{\alpha \partial_{\phi\phi}F}{\sin^2\theta_0} + \frac{\partial_{\theta\phi}F}{\sin\theta_0} \right) v \right].
\end{align*}
Here, we define
$$\beta\coloneq \frac{\gamma}{M_s(1+\alpha^2)}.$$
With this notation, the document rewrites the system in matrix form:
\[
\begin{pmatrix}
\dot{u}\\[4pt]
\dot{v}
\end{pmatrix}
=
\beta
\begin{pmatrix}
-\alpha \partial_{\theta\theta}F - \dfrac{\partial_{\theta\phi}F}{\sin\theta_0}
&
-\dfrac{\alpha \partial_{\theta\phi}F}{\sin\theta_0}
-
\dfrac{\partial_{\phi\phi}F}{\sin^2\theta_0}
\\[10pt]
\dfrac{\alpha \partial_{\theta\phi}F}{\sin\theta_0}
+ \partial_{\theta\theta}F
&
-\dfrac{\alpha \partial_{\phi\phi}F}{\sin^2\theta_0}
+
\dfrac{\partial_{\theta\phi}F}{\sin\theta_0}
\end{pmatrix}
\begin{pmatrix}
u\\[4pt]
v
\end{pmatrix}.
\]
Next, the matrix is decomposed and can be written as:
\begin{equation}
\begin{pmatrix}
\dot{u}\\[4pt]
\dot{v}
\end{pmatrix}
=
\beta (J - \alpha I) \mathcal{H}
\begin{pmatrix}
u\\[4pt]
v
\end{pmatrix}, \label{eq:first_dif_eq}
\end{equation}
where
\[
J =
\begin{pmatrix}
0 & -1\\
1 & 0
\end{pmatrix},
\qquad
I =
\begin{pmatrix}
1 & 0\\
0 & 1
\end{pmatrix},
\qquad
\mathcal{H} =
\begin{pmatrix}
\partial_{\theta\theta}F &
\dfrac{\partial_{\theta\phi}F}{\sin\theta_0}\\[8pt]
\dfrac{\partial_{\theta\phi}F}{\sin\theta_0} &
\dfrac{\partial_{\phi\phi}F}{\sin^2\theta_0}
\end{pmatrix}.
\]
Here, matrices $J$ and $I$ are respectively the matrix of $\pi/2$-rotation and identity matrix, and $\mathcal{H}$ is the \textit{Hessian matrix} of the free energy, composed of the second derivatives of the free energy density. It is important to note that we can equivalently write
$$\mathcal{H} =
\begin{pmatrix}
\partial_{uu}F &
{\partial_{uv}F}\\[8pt]
{\partial_{uv}F} &
{\partial_{vv}F}
\end{pmatrix},$$
where this time we express the second derivatives in the local coordinate system.

\subsection{Complex frequency and isotropic decay}

Equation \eqref{eq:first_dif_eq} defines a linear dynamical system whose
solutions are governed by the eigenvalues of $(J-\alpha I)\mathcal{H}$.
To present the immediate similarities with the damped harmonic oscillator, we investigate the second derivative of the equation. For an eigenmode of $(J-\alpha I)\mathcal{H}$, the second derivative of $\mathbf{v} = (u,v)^T$ satisfies
\begin{equation}
\ddot{\mathbf{v}} = \beta^2[(J-\alpha I)\mathcal{H}]^2 \,\mathbf{v}
= -\omega^2 \mathbf{v}.
\end{equation}

Here, $\omega$ is a complex frequency around the equilibrium, where its real part corresponds to an oscillation frequency, and its imaginary part corresponds to the damping. We note that although the local energy well may be anisotropic and the orbit elliptical, the linearized conservative dynamics is characterized by a single oscillation frequency.

It is not immediately obvious why the frequencies of oscillation and damping times along both principal axes should be the same. We present an intuitive argument following from the basic equation, which takes the following non-trivial form (the details of this derivation and the interpretation of the results are presented in the supporting information):
\begin{equation}
\begin{pmatrix}
u(t) \\[4pt]
v(t)
\end{pmatrix}
=
e^{\beta (J-\alpha I)\mathcal{H}t}
\begin{pmatrix}
u_0 \\[4pt]
v_0
\end{pmatrix}.
\end{equation}

The time evolution is therefore governed by a single matrix exponential. Its eigenvalues encode both the oscillatory part of the motion and the exponential relaxation toward equilibrium. To make an analogy with unidimensional systems, this is like having a sinusoidal term times a decaying exponential. 

Including damping produces a complex frequency
\[
\omega = a + ib,
\]
whose imaginary part determines the decay rate of the motion back to equilibrium.

Solving the characteristic equation for the eigenvalues of the matrix
$\beta(J-\alpha I)\mathcal{H}$ yields
\begin{align}
b &= \frac{\alpha\beta}{2}\,\mathrm{Tr}\,\mathcal{H},\\[4pt]
a &= \beta\sqrt{(1+\alpha^2)\det \mathcal{H} - \frac{\alpha^2}{4}(\mathrm{Tr}\,\mathcal{H})^2}.
\end{align}
Here $a$ is the oscillation frequency in the presence of damping and $b$ is
the exponential decay rate of the small–angle motion.

For the solution to make physical sense, we must have $b>0$, since a
negative $b$ would correspond to an unphysical exponential growth of the
magnetization amplitude.  This requirement is automatically satisfied for a stable equilibrium, for which the Hessian $\mathcal{H}$ is positive definite and hence
$\mathrm{Tr}\,\mathcal{H}>0$ and $\det \mathcal{H}>0$.

In the experimentally relevant regime $\alpha\ll 1$, the expressions above
simplify considerably.  Expanding in powers of $\alpha$ and retaining only
the leading terms gives
\begin{align}
\omega_{\mathrm{res}}
&\approx \beta\sqrt{\det \mathcal{H}},
\\[4pt]
\Delta\omega
&\approx \alpha\beta\,\mathrm{Tr}\,\mathcal{H},
\end{align}
where $\omega_{\mathrm{res}}$ denotes the resonance frequency (real part of
$\omega$) and $\Delta\omega$ is the full width at half maximum of the
resonance, proportional to the imaginary part.  These formulas show that, to lowest order in the damping parameter $\alpha$, the resonance frequency is controlled only by the geometric mean of the curvatures of the energy
well ($\sqrt{\det \mathcal{H}}$), whereas the linewidth is proportional to their sum
($\mathrm{Tr}\,\mathcal{H}$).

Physically, the first result reflects the fact that the conservative
(precessional) part of the dynamics is set by the local restoring torques,
which depend on the curvature of the free energy along both principal
directions around the minimum.  The second result shows that dissipation is
sensitive to how ``steep'' the well is when averaged over these directions:
a broader well (smaller $\mathrm{Tr}\,\mathcal{H}$) leads to a slower decay of oscillations, while a tighter well leads to faster damping, all else being
equal.

\subsection{Quality factor}
\label{subsec:Qfactor}

The resonance frequency and the linewidth naturally combine into the
dimensionless quality factor
\begin{equation}
Q
= \frac{\omega_{\mathrm{res}}}{\Delta\omega}
= \frac{\sqrt{\det \mathcal{H}}}{\alpha\,\mathrm{Tr}\,\mathcal{H}}.
\label{eq:Q_exact}
\end{equation}
Thus, the quality factor of small–angle precession is determined entirely by the Hessian of the free energy at the equilibrium point and by the Gilbert damping constant $\alpha$.  In particular, $Q$ is independent of the initial direction of the perturbation and of the coordinate system chosen to parametrize the unit sphere.

To make the dependence on the energy landscape more transparent, it is
convenient to express \eqref{eq:Q_exact} in terms of the eigenvalues
$\lambda_1$ and $\lambda_2$ of the Hessian $\mathcal{H}$.  Since $\mathcal{H}$ is symmetric and positive definite at a stable minimum, it can be diagonalized by an
orthonormal transformation, and we have
\[
\det \mathcal{H} = \lambda_1\lambda_2,
\qquad
\mathrm{Tr}\,\mathcal{H} = \lambda_1 + \lambda_2,
\]
so that
\begin{equation}
Q
= \frac{1}{\alpha}\,
\frac{\sqrt{\lambda_1\lambda_2}}{\lambda_1+\lambda_2}.
\label{eq:Q_lambda}
\end{equation}
It is useful to introduce the \emph{ellipticity} of the energy well,
defined as the ratio of eigenvalues
\[
e = \frac{\lambda_1}{\lambda_2} \quad (>0).
\]
Note that this definition is equivalent to the definition by the square of the ratio of the amplitudes of the oscillation in the principal axes. In terms of $e$, Eq.~\eqref{eq:Q_lambda} can be written as
\begin{equation}
Q
= \frac{1}{\alpha}
\frac{1}{\sqrt{e} + 1/\sqrt{e}}.
\label{eq:Q_ellipticity}
\end{equation}
This form shows that $Q$ depends only on the \emph{shape} of the well
through $e$, not on its overall scale: simultaneously increasing both
curvatures by the same factor leaves $Q$ unchanged, although it shifts the
absolute values of $\omega_{\mathrm{res}}$ and $\Delta\omega$.

The commonly used textbook expression
\[
Q_{\mathrm{usual}} = \frac{1}{2\alpha}
\]
is recovered from Eq.~\eqref{eq:Q_ellipticity} only in the special case of
an \emph{isotropic} minimum, where $\lambda_1=\lambda_2$ and hence $e=1$.
Indeed, setting $e=1$ in \eqref{eq:Q_ellipticity} gives
\[
Q(e=1)
= \frac{1}{\alpha}\,\frac{1}{1+1}
= \frac{1}{2\alpha}.
\]
In this case, the energy well is locally circular in the appropriate
orthonormal basis, and the precession is nearly circular; the standard
approximation is therefore accurate.

For anisotropic wells ($e\neq 1$), however, the quality factor deviates from
$1/(2\alpha)$.  The function
\[
f(e) = \frac{1}{\sqrt{e} + 1/\sqrt{e}}
\]
has a maximum at $e=1$. 
\begin{figure}[H]
    \centering
    \includegraphics[width=0.8\linewidth]{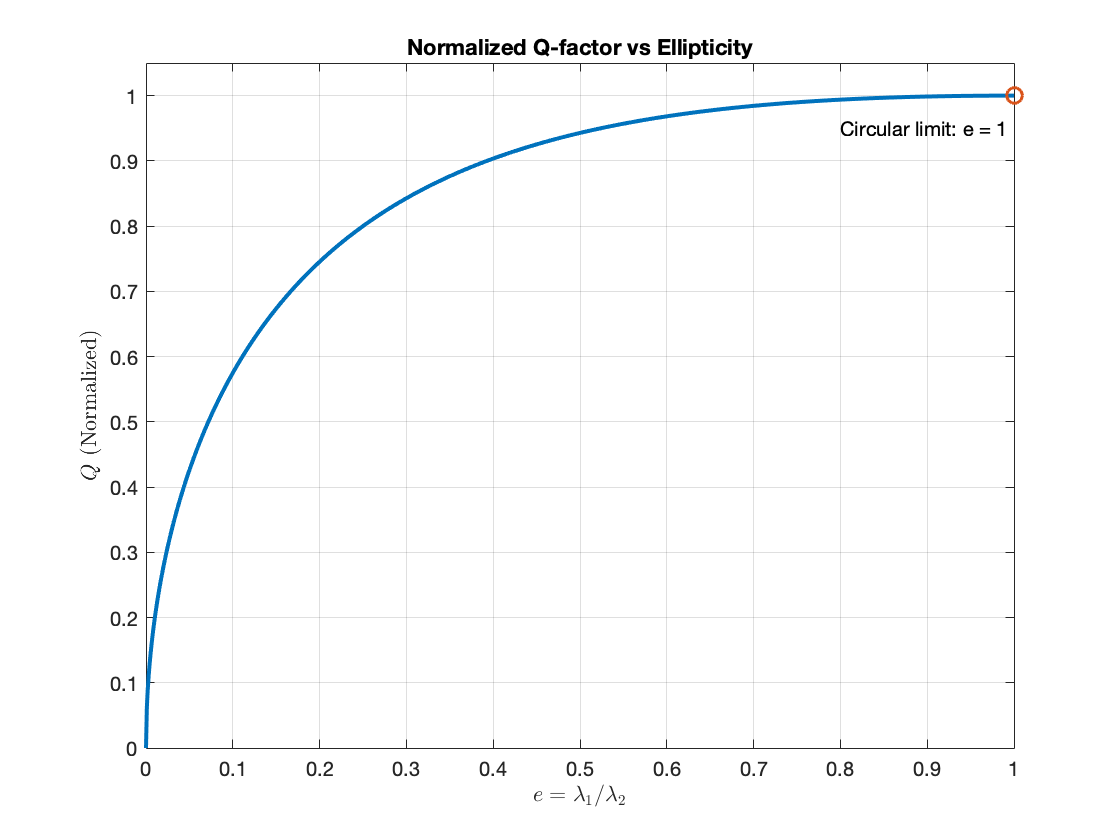}
    \caption{
        Quality factor $Q$ as a function of the ellipticity 
        $e = \lambda_1 / \lambda_2$ of the Hessian. 
    }
    \label{fig:Q_vs_e}
\end{figure}
Thus, 
\[
Q \le \frac{1}{2\alpha},
\]
with equality only for the circular case.  This means that the usual relation systematically \emph{overestimates} the coherence of the motion
whenever the energy minimum is elliptical.  In the extreme limit where one curvature tends to zero while the other remains finite (very elongated well), the product $\lambda_1\lambda_2$ tends to zero faster than the sum
$\lambda_1+\lambda_2$, so $Q\to 0$: the motion becomes overdamped and essentially non–oscillatory, in agreement with the behavior observed near
bifurcation curves where one of the eigenvalues of $\mathcal{H}$ vanishes.

From a physical perspective, Eq.~\eqref{eq:Q_exact} therefore provides a
precise criterion for the validity of the usual approximation
$Q=1/(2\alpha)$: it holds only when the curvature of the free–energy landscape is nearly isotropic in the neighborhood of the equilibrium.  In
more general situations, the anisotropy of $\mathcal{H}$ cannot be neglected, and the full Hessian–based expression must be used to correctly predict both
the resonance linewidth and the degree of coherence of magnetization
precession.

\section{Failure Regions for the Usual Value}
\label{sec:failure_regions}

In this section, we identify parameter regions where the approximation $Q_{\mathrm{usual}} \simeq 1/(2\alpha)$ overestimates the precession life time as the regions of bifurcation, i.e. the parameter boundaries where switching between single minimum and multiple minima occurs.

\subsection{Analytical example: symmetric cylinder}

We consider the case of magnetic shape anisotropy of a small ellipsoid described by the free energy density  
\begin{equation}
F(\mathbf m)
=
-\mu_0 M_s \mathbf H\cdot \mathbf m
+\frac12 \mu_0 M_s^2 \,\mathbf m^{\mathsf T} N \mathbf m,
\label{eq:F_zeeman_demag}
\end{equation}
where
\[
N=\mathrm{diag}(N_x,N_y,N_z)
\]
is the demagnetizing tensor, and $\mathbf m$ is the unit magnetization vector.

\label{subsec:symmetric_cylinder_bifurcation}

To illustrate analytically how the usual expression $Q=1/(2\alpha)$ breaks down near a {\it bifurcation}, we consider a symmetric cylinder with a demagnetizing tensor
\[
\mathbf{N}=\mathrm{diag}\!\left(\frac12,\,0,\,\frac12\right),
\]
and an external magnetic field applied along the symmetry axis:
\[
\mathbf{H}=H\,\hat{\mathbf z}, \qquad h=\frac{H}{M_s}.
\]
We retain only the Zeeman and demagnetizing contributions to the free-energy density:
\begin{equation}
F(\mathbf m)
=
-\mu_0 M_s H m_z
+\frac12 \mu_0 M_s^2
\left(
\frac12 m_x^2+\frac12 m_z^2
\right).
\label{eq:F_symcyl_Cartesian}
\end{equation}
Introducing the reduced energy
\begin{equation}
f(\mathbf m)=\frac{F(\mathbf m)}{\mu_0 M_s^2}=
-h\,m_z+\frac14\left(m_x^2+m_z^2\right)\,\label{eq:f_symcyl_Cartesian}
\end{equation}
and using spherical coordinates, we obtain
\begin{equation*}
f(\theta,\phi)
=
-h\cos\theta
+\frac14\left(\sin^2\theta\cos^2\phi+\cos^2\theta\right).
\label{eq:f_symcyl_spherical}
\end{equation*}
It is convenient to rewrite this as
\begin{equation*}
f(\theta,\phi)
=
-h\cos\theta
+\frac14
-\frac14\sin^2\theta\,\sin^2\phi.
\label{eq:f_symcyl_simplified}
\end{equation*}
The first derivatives are
\begin{align}
\partial_\theta f
&=
h\sin\theta-\frac12\sin\theta\cos\theta\,\sin^2\phi,
\label{eq:dthetaf_symcyl}
\\[4pt]
\partial_\phi f
&=
-\frac12\sin^2\theta\,\sin\phi\cos\phi.
\label{eq:dphif_symcyl}
\end{align}
Stationary points satisfy $\nabla f = 0$. From \eqref{eq:dphif_symcyl}, we have
\[
\sin^2\theta\,\sin\phi\cos\phi=0,
\]
so the possible cases are:
\[
\sin\theta=0,
\qquad
\sin\phi=0,
\qquad
\cos\phi=0.
\]
First two cases correspond to trivial extremum locations which are not interesting to us. Solutions with $\cos\phi=0$ give
\[
\phi=\frac{\pi}{2},\ \frac{3\pi}{2}.
\]
Substituting into \eqref{eq:dthetaf_symcyl} yields
\[
\sin\theta\left(h-\frac12\cos\theta\right)=0.
\]
Besides the polar solutions, we obtain the nontrivial equilibria
\begin{equation*}
\cos\theta=2h.
\label{eq:cos_theta_2h}
\end{equation*}
These exist only when
\begin{equation*}
|2h|\le 1
\quad\Longleftrightarrow\quad
|h|\le \frac12.
\label{eq:existence_condition_symcyl}
\end{equation*}
Therefore, for $0<h<1/2$, the system possesses two symmetry-related off-axis equilibria,
\begin{equation}
(\theta_0,\phi_0)=
\left(\arccos(2h),\frac{\pi}{2}\right),
\qquad
\left(\arccos(2h),\frac{3\pi}{2}\right),
\label{eq:two_minima_symcyl}
\end{equation}
which correspond to two minima of the free-energy landscape. At $h=1/2$, these two minima merge with the north pole, and for $h>1/2$ only the north-pole minimum remains. Thus the bifurcation occurs at
\begin{equation}
h_c=\frac12.
\label{eq:hc_symcyl}
\end{equation}

To characterize the bifurcation, we compute the Hessian of the reduced energy in the local tangent-plane coordinates. The second derivatives are
\begin{align*}
\partial_{\theta\theta}f
&=
h\cos\theta-\frac12(\cos^2\theta-\sin^2\theta)\sin^2\phi,
\\[4pt]
\partial_{\phi\phi}f
&=
-\frac12\sin^2\theta(\cos^2\phi-\sin^2\phi),
\\[4pt]
\partial_{\theta\phi}f
&=
-\sin\theta\cos\theta\sin\phi\cos\phi.
\end{align*}
At the equilibrium point \eqref{eq:two_minima_symcyl}, where
\[
\phi_0=\frac{\pi}{2}\ \text{or}\ \frac{3\pi}{2},
\qquad
\cos\theta_0=2h,
\qquad
\sin^2\theta_0=1-4h^2,
\]
we obtain
\begin{align*}
\partial_{\theta\phi}f(\theta_0,\phi_0)&=0,
\\
\partial_{\theta\theta}f(\theta_0,\phi_0)
&=
\frac{1-4h^2}{2},
\\
\frac{1}{\sin^2\theta_0}\partial_{\phi\phi}f(\theta_0,\phi_0)
&=
\frac12.
\end{align*}
Hence the local Hessian (of the reduced energy) in the coordinates $(u,v)$ is diagonal:
\begin{equation}
\mathcal H_f
=
\begin{pmatrix}
\dfrac{1-4h^2}{2} & 0 \\[6pt]
0 & \dfrac12
\end{pmatrix}.
\label{eq:Hessian_symcyl}
\end{equation}
Its eigenvalues are therefore
\begin{equation}
\lambda_1=\frac{1-4h^2}{2},
\qquad
\lambda_2=\frac12.
\label{eq:eigenvalues_symcyl}
\end{equation}
This result makes the bifurcation transparent: as $h\to 1/2^{-}$,
\[
\lambda_1\to 0,
\qquad
\lambda_2\to \frac12,
\]
so one principal curvature of the free-energy minimum collapses while the other remains finite. The minimum becomes extremely elongated, and the precession loses its circular character. Using the general formulas derived above,
\begin{equation}
\omega_{\mathrm{res}}
\propto
\sqrt{\lambda_1\lambda_2},
\qquad
\Delta\omega
\propto
\alpha(\lambda_1+\lambda_2),
\qquad
Q=
\frac{1}{\alpha}
\frac{\sqrt{\lambda_1\lambda_2}}{\lambda_1+\lambda_2},
\end{equation}
we obtain for the symmetric cylinder
\begin{align}
\omega_{\mathrm{res}}
&\propto
\frac12\sqrt{1-4h^2},
\\[4pt]
\Delta\omega
&\propto
\alpha(1-2h^2),
\\[4pt]
Q
&=
\frac{1}{2\alpha}\,
\frac{\sqrt{1-4h^2}}{1-2h^2}.
\label{eq:Q_symcyl_analytic}
\end{align}
Therefore,
\begin{equation}
Q\to 0
\qquad\text{as}\qquad
h\to \frac12^{-},
\end{equation}
even though $\alpha$ remains fixed and small. This shows explicitly that the isotropic estimate
\[
Q=\frac{1}{2\alpha}
\]
fails near the bifurcation point. Analytical treatment presented above collinear is valid for collinear geometry. For noncollinear geometries calculations can be performed numerically.

\subsection{Computational example: arbitrary ellipsoid}

We now present a computational demonstration of the relation between the bifurcation boundary and the $Q$-factor for a non-trivial geometry. We choose an arbitrary geometry for the nanomagnet, characterized by the demagnetizing tensor
\[
\mathbf{N}=\mathrm{diag}\!\left(0.3,\,0.05,\,0.65\right).
\]
We set the external magnetic field on the $x-z$ plane. We note the external field angle $\xi$ as the angle between the $z$-axis and the external field. For different parameter configurations $(h, \xi)$, we numerically compute the Hessian matrix projection at the minimum location. Using that, we compute the values of the $Q$-factor, normalize, and plot it on the $(h, \xi)$ using a color map. On a separate program which runs independently from the first one, we compute the number of extrema (thus the minima) for each $(h, \xi)$, and determine the parameter boundary where this number switches. Overlaying this boundary on the computed map of the $Q$-factors reveals the expected matching between the boundary and the region where we have $Q\rightarrow 0$. Note that we do not compute the $Q$-factors for the two minima region. Because of the asymmetry of our geometry, the ellipticities and the related factors will differ between the two minima. However, it is enough to see the decrease of the $Q$-factor for the single minimum region to see its overlap with the bifurcation boundary.

\begin{figure}
    \centering
    \includegraphics[width=1.0\linewidth]{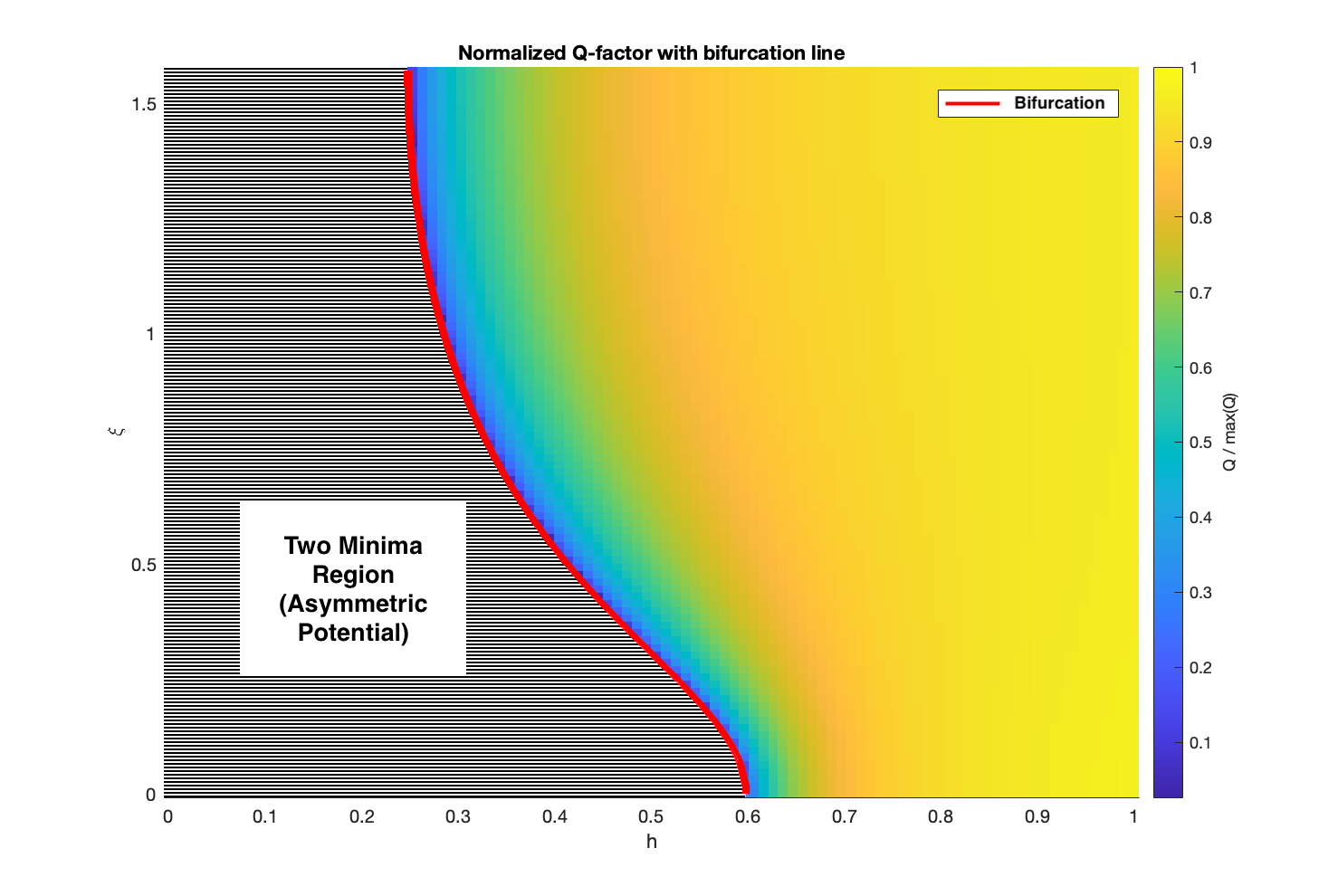}
    \caption{Overlay of the bifurcation boundary on the $Q$-factor plot}
    \label{fig:bifurcation}
\end{figure}

\section{Summary and Conclusions}

In this work, we have developed a compact and geometry-agnostic formulation of small-angle 
magnetization dynamics within the Landau--Lifshitz--Gilbert equation by expressing the local restoring torques in terms of the Hessian of the free energy at equilibrium. This approach shows that the precession frequency and the damping rate depend on distinct invariants of the Hessian: 
the resonance frequency scales as $\omega_{\mathrm{res}} \propto \sqrt{\det \mathcal{H}}$, while the linewidth is 
proportional to $\Delta\omega \propto \mathrm{Tr}\,\mathcal{H}$.

A direct consequence of this separation is that the commonly used approximation 
$Q = 1/(2\alpha)$ represents the maximum attainable quality factor, achieved only for circular energy minima with $\lambda_1 = \lambda_2$. In generic situations, including those involving shape or magneto-crystalline anisotropies, notably in the vicinity of bifurcations, the eigenvalues of the Hessians become highly unequal. This leads to a collapse of $Q$ and to the appearance of overdamped magnetization dynamics even when the Gilbert damping parameter $\alpha$ is small. Numerical simulations of the LLG equation for planar, symmetric, and asymmetric geometries confirm these predictions and demonstrate the universality of the Hessian-based approach.

The theoretical framework presented here provides a minimal and broadly applicable tool for predicting linewidth, damping, and precessional behavior across diverse magnetic configurations. It depends only on the local curvature of the free energy and can be extended to more complex anisotropies, multilayer structures, or dynamical regimes beyond the small-angle approximation. Future work could explore the application of this formalism to systems with time-dependent fields \cite{alvarez2000quasiperiodicity} and  parametric driving \cite{alekhin2023quantitative} as well as the experimental determination of the Gilbert damping parameter \cite{salikhov2019gilbert}.


\begin{acknowledgments}
The support of the Physics Department of \'Ecole Polytechnique and Institut Polytechnique de Paris within the framework of a {\it Projet de Recherche en Laboratoire} is acknowledged.
\end{acknowledgments}

\bibliography{report}

\begin{thebibliography}{10}

\bibitem{farle1998ferromagnetic}
Farle, M., ``Ferromagnetic resonance of ultrathin metallic layers,'' {\em Reports on progress in physics}~{\bf 61}(7),  755 (1998).

\bibitem{Landau1935}
Landau, L.~D. and Lifshitz, L.~M., ``On the theory of the dispersion of magnetic permeability in ferromagnetic bodies,'' {\em Physik. Zeits. Sowjetunion}~{\bf 8},  153--169 (1935).

\bibitem{lakshmanan2011fascinating}
Lakshmanan, M., ``The fascinating world of the landau--lifshitz--gilbert equation: an overview,'' {\em Philosophical Transactions of the Royal Society A: Mathematical, Physical and Engineering Sciences}~{\bf 369}(1939),  1280--1300 (2011).

\bibitem{salikhov2019gilbert}
Salikhov, R., Alekhin, A., Parpiiev, T., Pezeril, T., Makarov, D., Abrudan, R., Meckenstock, R., Radu, F., Farle, M., Zabel, H., et~al., ``Gilbert damping in nifegd compounds: Ferromagnetic resonance versus time-resolved spectroscopy,'' {\em Physical Review B}~{\bf 99}(10),  104412 (2019).

\bibitem{BesseJMMM}
Besse, V., Golov, A.~V., Vlasov, V.~S., Alekhin, A., Kuzmin, D., Bychkov, I.~V., Kotov, L.~N., and Temnov, V.~V., ``Generation of exchange magnons in thin ferromagnetic films by ultrashort acoustic pulses,'' {\em J. Magn. Magn. Mater.}~{\bf 502},  166320 (2020).

\bibitem{ghita2023anatomy}
Ghita, A., Mocioi, T.-G., Lomonosov, A.~M., Kim, J., Kovalenko, O., Vavassori, P., and Temnov, V.~V., ``Anatomy of ultrafast quantitative magnetoacoustics in freestanding nickel thin films,'' {\em Physical Review B}~{\bf 107}(13),  134419 (2023).

\bibitem{gurevich2020magnetization}
Gurevich, A.~G. and Melkov, G.~A.,  [{\em Magnetization oscillations and waves}{\nolinebreak\hspace{0.1em}]}, CRC press (2020).

\bibitem{alekhin2023quantitative}
Alekhin, A., Lomonosov, A.~M., Leo, N., Ludwig, M., Vlasov, V.~S., Kotov, L., Leitenstorfer, A., Gaal, P., Vavassori, P., and Temnov, V., ``Quantitative ultrafast magnetoacoustics at magnetic metasurfaces,'' {\em Nano Letters}~{\bf 23}(20),  9295--9302 (2023).

\bibitem{Osborn1945}
Osborn, J.~A., ``Demagnetizing factors of the general ellipsoid,'' {\em Phys. Rev.}~{\bf 67}(11 and 12),  351 (1945).

\bibitem{alvarez2000quasiperiodicity}
Alvarez, L.~F., Pla, O., and Chubykalo, O., ``Quasiperiodicity, bistability, and chaos in the landau-lifshitz equation,'' {\em Physical Review B}~{\bf 61}(17),  11613 (2000).

\end{thebibliography}
\bibliographystyle{spiebib} 

\end{document}